\documentclass[onecolumn,12pt]{article}
\usepackage{amssymb}
\usepackage{graphicx}
\usepackage{graphics}
\usepackage{enumerate}
\usepackage{psfig}
\usepackage{geometry}
\begin{document}

\title{Quantum Noise Randomized Ciphers}
\author{Ranjith Nair\thanks{Email: nair@eecs.northwestern.edu}, Horace P. Yuen, Eric Corndorf,  Takami Eguchi, and Prem Kumar \\ Center for Photonic Communication and Computing\\
Department of Electrical Engineering and Computer Science\\
Northwestern University, Evanston, IL 60208} \maketitle

PACS: 03.67.Hk, 42.50.Ar
\newline
\begin{abstract}
We review the notion of a classical random cipher and its
advantages. We sharpen the usual description of random ciphers to
a particular mathematical characterization suggested by the
salient feature responsible for their increased security. We
describe a concrete system known as $\alpha\eta$ and show that it
is equivalent to a random cipher in which the required
randomization is effected by coherent-state quantum noise. We
describe the currently known security features of $\alpha\eta$ and
similar systems, including lower bounds on the unicity distances
against ciphertext-only and known-plaintext attacks. We show how
$\alpha\eta$ used in conjunction with any standard stream cipher
such as AES (Advanced Encryption Standard) provides an additional,
qualitatively different layer of security from physical encryption
against known-plaintext attacks on the key.
We
refute some claims in the literature 
that $\alpha\eta$ is equivalent to a non-random stream cipher.
\end{abstract}

\section{Introduction}
The possibility of achieving greater secrecy by introducing
additional randomness into the plaintext of a cipher before
encryption was known, according to \cite{massey88}, already to
Gauss, in the form of the so-called `homophonic substitution'.
Such a procedure is an example of a \emph{random cipher}
\cite{massey88,yuen05qph}. The advantage of a random cipher not
present in standard nonrandom ciphers is that it can provide
information-theoretic security of the key against statistical
attacks, and possibly known-plaintext attacks (See Appendix A and
also \cite{yuen05qph}). A somewhat detailed description of these
possibilities is one of the goals of this paper. In spite of the
potential advantages of random ciphers, a large obstacle in their
deployment is the bandwidth expansion, or more accurately data
rate reduction, that is needed to operate all previous random
ciphers. Also, it is not currently possible to generate true
random numbers at speeds high enough for random ciphers to operate
at sufficiently high data rates ($\sim$ Mbps is the current upper
limit for random number generation). The quantum noise in optical
coherent-state signals may be utilized for this purpose, and
quantum optical effects seem to be the only technologically
feasible way to generate $>$ Gbps true random numbers.  A
particular quantum noise-based random cipher, called $\alpha\eta$,
that also does not entail data rate reduction, has already been
proposed and implemented \cite{barbosa03,yuen03} at Northwestern
University. In a previous preprint \cite{yuen05qph}, $\alpha\eta$
was discussed  concomitantly with that of the closely related key
generation system called $\alpha\eta$-KG.  Since the features of
$\alpha\eta$ direct encryption are subtle and complex enough, we
take the approach in this paper of discussing just the
$\alpha\eta$ encryption system in its own right, and analyze
quantitatively its random cipher feature. Doing so will hopefully
also avert many possible confusions with $\alpha\eta$-Key
Generation, such as those in \cite{nishioka04,nishioka05}.  In
particular, we will set up in detail the proper framework to
understand and analyze the security issues involved. Note that the
present paper can be understood independently of ref.
\cite{yuen05qph}, the relevant terminology and results from which
are summarized in Section 2.1 and Appendix A of this paper.

Following our discussion of random ciphers in general and the
$\alpha\eta$ cryptosystem, we show that $\alpha\eta$ security is
equivalent to that of a corresponding classical random cipher. We
show how quantum noise allows some degree of randomization in
$\alpha\eta$ without sacrificing data rate, and quantify the
randomization by two different parameters corresponding to
ciphertext-only and known-plaintext attacks. We also show how
$\alpha\eta$ can be operated on top of a standard cipher like AES
to provide additional, qualitatively different, security based on
quantum noise against known-plaintext attacks on the key. However,
information-theoretically, ciphertext-only attack on the key is
possible with the original $\alpha\eta$. We will indicate what
additional techniques can alleviate this problem, without going
into any detailed analysis to be presented at a later time.
Generally, only search-complexity based security will be
quantitatively described in this paper. Finally, we rebut the
claims in \cite{nishioka04,nishioka05} that $\alpha\eta$ security
is equivalent to that of a standard stream cipher and that
$\alpha\eta$ is nonrandom.

The plan of this paper is as follows: In Section 2, we provide the
necessary review of standard cryptography. In addition, we define
the random cipher concept quantitatively and point out the
available results on random cipher security. This sets the stage
for our definitions in Section 3 that characterize a \emph{quantum
cipher} and a \emph{quantum random cipher}, which are both ciphers
in which the ciphertext is in the form of a quantum state. In
Section 4, we describe the $\alpha\eta$ system in detail, show its
quantum random cipher characteristics, and highlight its
advantages. In Section 5, we respond to the criticisms on
$\alpha\eta$ made by Nishioka et al \cite{nishioka04,nishioka05} in a
further elaboration of the quantitative random cipher character of $\alpha\eta$.

\section{Standard Cryptography and Random Ciphers}

\subsection{Standard Symmetric-Key Cryptography}
We review the basics of symmetric-key data encryption. Further
details can be found in, e.g., \cite{massey88,stinson}. Throughout
the paper, random variables will be denoted by \emph{upper-case}
letters such as $K,X_1$ etc.  It is sometimes necessary to
consider explicitly sequences of random variables $(X_1,X_2,
\ldots, X_n)$. We will denote such \emph{vector} random variables
by a \emph{boldface} upper-case letter $\mathbf{X}_n$ and,
whenever necessary, indicate the length of the vector ($n$ in this
case) as a subscript. Confusion with the $n$-th component $X_n$ of
$\mathbf{X}_n$ should not arise as the latter is a boldface
vector. Particular values taken by these random variables will be
denoted by similar \emph{lower-case} alphabets. Thus, particular
values taken by the key random variable $K$ are denoted by $k, k'$
etc. Similarly, a particular value of $\mathbf{X}_n$ can be
denoted $\mathbf{x}_n$. The plaintext alphabet will be denoted
$\mathcal{X}$, the set of possible key values $\mathcal{K}$ and
the ciphertext alphabet $\mathcal{Y}$. Thus, for example, the
sequences $\mathbf{x}_n \in \mathcal{X}^n$. In most nonrandom
ciphers, $\mathcal{X}$ is simply the set $\{0,1\}$ and
$\mathcal{Y}=\mathcal{X}$.

With the above notations, the $n$-symbol long \emph{plaintext}
(i.e., the message sequence that needs to be encrypted) is denoted
by the random vector $\mathbf{X}_n$, the \emph{ciphertext} (i.e.,
the output of the encryption mechanism) is denoted by
$\mathbf{Y}_n$ and the secret key used for encryption is denoted
by $K$. In this paper, we will often call the legitimate sender of
the message `Alice', the legitimate receiver `Bob', and the
attacker (or eavesdropper) `Eve'. Note that although the secret
key is typically a sequence of bits, we do not use vector notation
for it since the bits constituting the key will not need to be
singled out separately in our considerations in this paper. In
standard cryptography, one usually deals with \textit{nonrandom
ciphers}. These are ciphers for which the ciphertext is a function
of only the plaintext and key. In other words, there is an
encyption function $E_{k}(\cdot)$ such that:
\begin{equation} \label{encryption}
\mathbf{y}_n=E_{k}(\mathbf{x}_n). \end{equation} There is a
corresponding decryption function $D_{k}(\cdot)$ such that:
\begin{equation} \label{decryption}
\mathbf{x}_n=D_{k}(\mathbf{y}_n). \end{equation} In such a case,
the $X_i$ and $Y_i, i=1,\ldots, n$ are usually taken to be
from the same alphabet. 

In contrast, a \emph{random cipher} makes use of an additional
random variable $R$ called the \emph{private randomizer}
\cite{massey88}, generated by  Alice while encrypting the
plaintext and known only to her, if at all. Thus the ciphertext is
determined as follows:
\begin{equation} \label{randomcipher}
\mathbf{y}_n=E_{k}(\mathbf{x}_n, r).
\end{equation}
Because of the additional randomness in the ciphertext, it
typically happens that the ciphertext alphabet $\mathcal{Y}$ needs
to be larger than the plaintext alphabet $\mathcal{X}$ (or else,
$\mathbf{Y}$ is a longer sequence than $\mathbf{X}$, as in
homophonic substitution). It may even be a continuous infinite
alphabet, e.g. an analog voltage value. However, we still require,
as in \cite{massey88}, that Bob be able to decrypt with just the
ciphertext and key (i.e., without knowing $R$), so that there
exists a function $D_{k}(\cdot)$ such that Eq.(\ref{decryption})
holds. We note that random ciphers are called `privately
randomized ciphers' in Ref. \cite{massey88} -- we will however use
the shorter term `random cipher' (Note that `random cipher' is
used in a completely different sense by Shannon \cite{shannon49}).

We note  that the presence or absence of the private randomizer
$R$ may be indicated using the conditional Shannon entropy (We
assume a basic familiarity with Shannon entropy and conditional
entropy. See any information theory textbook, e.g.,
\cite{cover91}.). For nonrandom ciphers, we have from
Eq.(\ref{encryption}) that
\begin{equation} \label{nonrandom}
H(\mathbf{Y}_n|K \mathbf{X}_n)=0. \end{equation}   On the other
hand, a \textit{random cipher} satisfies
\begin{equation} \label{random}
H(\mathbf{Y}_n|K\mathbf{X}_n) \neq 0, \end{equation} due to the
randomness supplied by the private randomizer $R$. The decryption
condition Eqs.(\ref{decryption}) for both random and nonrandom
ciphers has the entropic characterization:
\begin{equation} \label{decrypt} H(\mathbf{X}_n|K\mathbf{Y}_n)=0.
\end{equation}
Note that this characterization of a random cipher is problematic
when the ciphertext alphabet is continuous, as could be the case
with $\alpha\eta$, because then the Shannon entropy is not
defined. It may be argued that the finite precision of measurement
forces the ciphertext alphabet to be discrete. Indeed, in
Sec.~2.2, we define a parameter $\Lambda$ that characterizes the
``degree of randomness'' of a random cipher. In any case, the
definition makes sense, similar to Eq.~(\ref{random}), only when
the ciphertext alphabet is finite, or at most discrete.

In the cryptography literature, the characterization of a general
random cipher is limited to that given by Eqs.
(\ref{randomcipher}) and (\ref{random}). See, e.g.,
\cite{massey88}. In the next section, we will see that the
purposes of cryptographic security suggest a sharper quantitative
definition of a random cipher involving a pertinent security
parameter $\Gamma$. This new definition, unlike (\ref{random}),
will be meaningful irrespective of whether the ciphertext alphabet
is discrete or continuous. Before we discuss the above new
definition of random ciphers, we conclude this section with some
important cryptographic terminology.


By \textit{standard cryptography}, we shall mean that Eve and Bob
both observe the same ciphertext random variable, i.e.,
$\mathbf{Y}^{\rm E}_{n}=\mathbf{Y}_{n}^{\rm B}=\mathbf{Y}_{n}$.
Thus, standard cryptography includes usual mathematical
private-key (and also public-key) cryptography but excludes
quantum cryptography and classical-noise cryptography
\cite{maurer93}. For a standard cipher, random or nonrandom, one
can readily prove from the above definitions the following result
known as the \emph{Shannon limit} \cite{massey88,shannon49}:
\begin{equation} \label{shannonlimit} H(\mathbf{X}_n|\mathbf{Y}_n) \leq H(K).
\end{equation}
This result may be thought of as saying that no matter how long
the plaintext sequence is, the attacker's uncertainty on it
\emph{given the ciphertext} cannot be greater than that of the
key. This condition is of crucial importance in both direct
encryption and key generation, as brought out in refs.
\cite{yuen03,yuen05qph,yuen05pla,pra05,yuan05}, but was missed in
previous criticisms of $\alpha\eta$
\cite{nishioka04,nishioka05,loko05}.

By \textit{information-theoretic security} (or \emph{IT security})
on the data, we mean that Eve cannot, even with unlimited
computational power, pin down uniquely the plaintext from the
ciphertext, i.e.,
\begin{equation}\label{ITsecurity}
H(\mathbf{X}_n|\mathbf{Y}_n)\neq 0.
\end{equation}
The level of such security may be quantified by
$H(\mathbf{X}_n|\mathbf{Y}_n)$. Shannon has defined
\textit{perfect security} \cite{shannon49} to mean that the
plaintext is statistically independent of the ciphertext, i.e.,
\begin{equation} H(\mathbf{X}_n|\mathbf{Y}_n)=H(\mathbf{X}_n).
\end{equation}
With the advent of quantum cryptography, the term `unconditional
security' has come to be used, unfortunately in many possible
senses. By \emph{unconditional security}, we shall mean
near-perfect information-theoretic security against all attacks
consistent with the known laws of quantum physics.

Incidentally, note that the Shannon limit Eq.~(\ref{shannonlimit})
immediately shows that perfect security can be attained only if
$H(\mathbf{X}_n)\leq H(K)$, so that, in general, the key needs to
be as long as the plaintext.


\subsection{Random Ciphers -- Quantitative Definition}
As mentioned in the previous section, the characterization of a
general random cipher merely using Eq.~(\ref{randomcipher}) or
(\ref{random}) is perhaps not well-motivated. The reason for
studying random ciphers is in fact  the belief that they enhance
the security of the cipher against various attacks. By bringing
into focus the intuitive mechanism by which a random cipher may
provide greater security than a nonrandom counterpart against
known-plaintext attacks, we will propose one possible quantitative
characterization of a general random cipher (or more exactly, a
general random \emph{stream} cipher. See below.). For a
description of known-plaintext and other attacks on ciphers,
together with the known results on their security, we refer the
reader to Appendix A.

We now discuss the intuitive mechanism of security enhancement in
a random cipher. To this end, a schematic depiction of encryption
and decryption with a random cipher is given in Fig.~1. For a
binary alphabet $\mathcal{X}=\{0,1\}$, let
$\mathcal{X}^n=\{a_1,\ldots, a_N\}$ be the set of $N=2^n$ possible
plaintext $n$-sequences. Let $k$ be a particular key value. One
can view the key $k$ as dividing the ciphertext space
$\mathcal{Y}^n$ into $N$ parts, denoted by the
$\mathcal{A}_{a_j}^k, j \in \{1, \ldots, N\},$ in the figure.
Encryption of plaintext $a_j$ proceeds by first determining the
relevant region $\mathcal{A}_{a_j}^k$ and randomly selecting (this
is the function of the private randomizer) as ciphertext some
$y\in\mathcal{A}_{a_j}^k$. The decryption condition
Eq.(\ref{decryption}) is satisfied by virtue of the regions
$\mathcal{A}_{a_j}^k$ being disjoint for a given $k$. Also shown
in Fig. 1 is the situation where a different key value $k'$ is
used in the system. The associated partition of $\mathcal{Y}^n$
consists of the sets $\mathcal{A'}_{a_j}^{k}$ that are shown with
shaded boundaries in Fig. 1. The \emph{important point} here is
that the respective partitions of the ciphertext space for the key
values $k$ and $k'$ should be sufficiently `intermixed'. More
precisely, for any given plaintext $a_j$, and any observed
ciphertext $\mathbf{y}_n$, we require that there exist
sufficiently many key values $k$ (and hence a sufficiently large
probability of the set of possible keys corresponding to a given
plaintext and observed ciphertext) for which $\mathbf{y}_n \in
\mathcal{A}_{a_j}^k$. In other words, a given plaintext-ciphertext
pair can be connected by many possible keys. This is the intuitive
basis why random ciphers offer better quantitative security (as
measured either by Eve's information on the key or her complexity
in finding it; see Sec. 4.2-4.4 for a discussion of $\alpha\eta$
security) than nonrandom ciphers against known-plaintext attacks.

\begin{figure} [htbp]
\begin{center}
\rotatebox{-90} {
\includegraphics[scale=0.5]{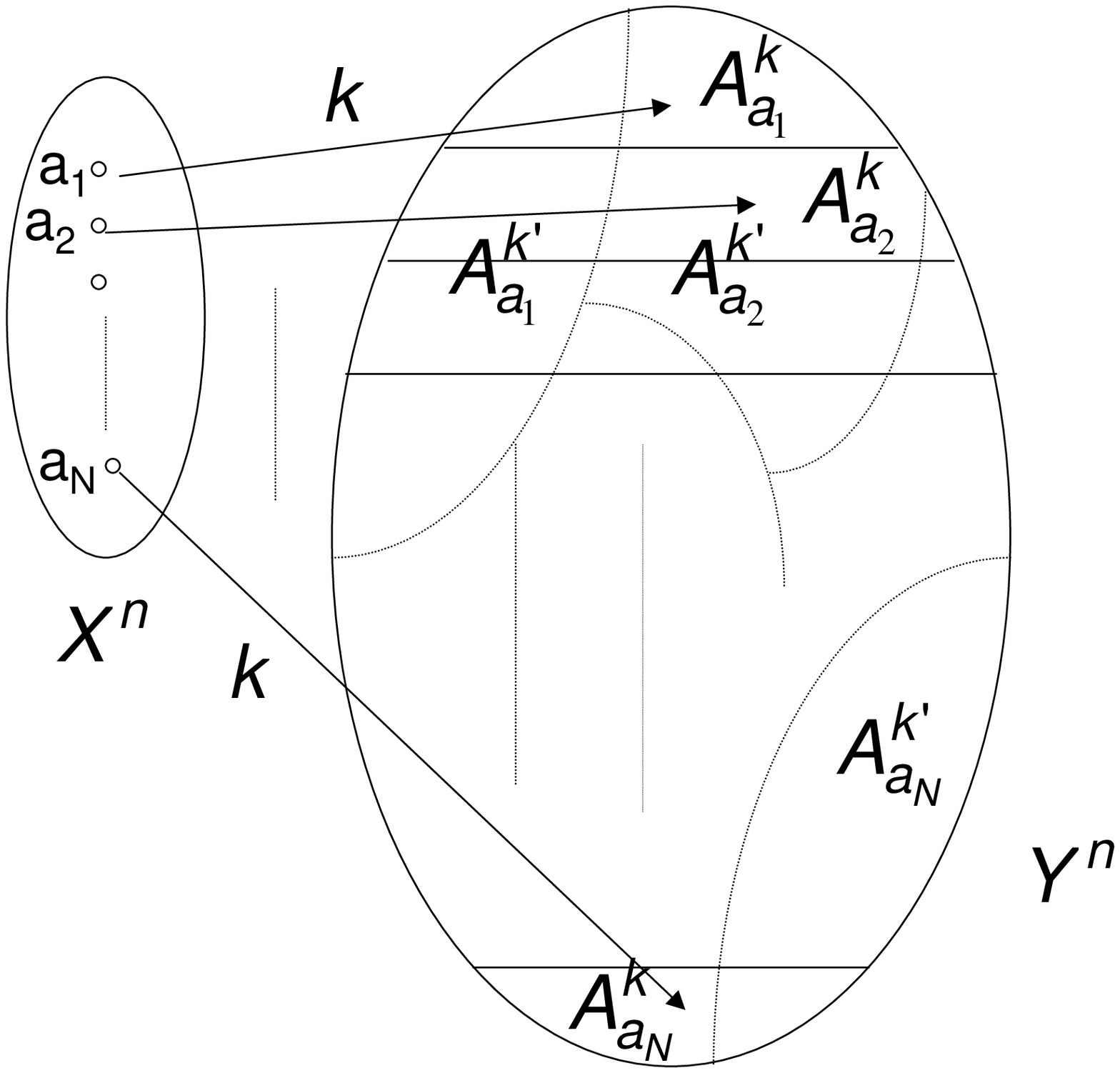}}
\caption{Schematic of a random cipher: The plaintexts $a_i$ are
carried, under the key $k$, into the corresponding regions
$A_{a_j}^k$ of ciphertext space $Y^n$. The subsets of $Y^n$
associated with a different key value $k'$ are shown with curved
boundaries.}
\end{center}
\end{figure}

While the above arguments hold for any type or random cipher
whatsoever, we will restrict our scope to the so-called
\emph{stream ciphers}. Most ciphers in current use (which are all
nonrandom), such as AES, are stream ciphers \cite{stinson}. In a
nonrandom stream cipher, the key $K$ is first expanded using a
deterministic function into a much longer sequence $(Z_1,\ldots,
Z_n)$ called the \emph{keystream} or \emph{running key}. The
defining property of a \emph{stream cipher} is that the $i$-th
ciphertext symbol $y_i$ be a function of just the $i$-th keystream
symbol $z_i$ and the earlier and current plaintext symbols
$x_1,\ldots,x_i$:
\begin{equation} \label{streamcipher}
y_i=E^i(x_1, \ldots, x_i;z_i). \end{equation} It follows that
decryption of the first $i$ symbols of plaintext is possible from
the first $i$ symbols of ciphertext and the running key. A
\emph{synchronous} stream cipher is one for which
\begin{equation} \label{syncstreamcipher} y_i=E^i(x_i;z_i).
\end{equation} Thus, the $i$-th ciphertext symbol depends only on
the $i$-th plaintext symbol and the $i$-th keystream symbol, i.e.,
the cipher is memoryless. For our discussion of random ciphers, we
will  restrict ourselves for concreteness to the case of
\emph{random stream ciphers}, that are defined by:
\begin{equation} \label{randomstreamcipher}
y_i=E^i(x_1, \ldots, x_i; z_i; r_i). \end{equation} Here, the
$\{R_i\}$ are randomizers that may be assumed to be independent
random variables (this is the case in $\alpha\eta$), but this is
not necessary. In the rest of the paper, a \emph{random cipher}
will always mean a \emph{random stream cipher}.

 For a nonrandom stream
cipher given by Eq.~(\ref{streamcipher}), it is usually the case
that given the plaintext vector $\mathbf{x}_i$ of length $i$ and
ciphertext symbol $y_i$, the value of the keystream $z_i$ is
uniquely determined. This is typically the case also in a random
stream cipher \emph{when the value $r$ taken by the randomizer
$R_i$ is known}. In the absence of such knowledge, however, the
different possible values taken by $R_i$ will in general allow
many different values of the keystream for the given plaintext
vector and ciphertext symbol. The more such possibilities exist,
the less information is obtained about the keystream and the more
`secure' the cipher is. Our quantitative definition of random
cipher given below introduces a parameter $\Gamma$ that provides
one way of quantifying the different knowledge of the keystream
obtained in the above two scenarios by the number of additional
possible keystreams for a given pair of input data and
corresponding ciphertext symbols.
\\ \\
\textbf{Definition} (\emph{$\Gamma$- Random Cipher})
\textbf{:}\\
A $\Gamma$-Random Cipher is a random stream cipher of the form of
Eq.~(7) for which the following condition holds: \\For every
plaintext sequence, $\mathbf{x}_i$, for every $i$, for every
ciphertext symbol $y_i$ obtainable by encryption of
$\mathbf{x}_i$, and for every value $r$ of $R_i$,
\begin{equation} \label{Gammarandomcipherdef}
|\{z_i | y_i=E^i(x_1, \ldots, x_i; z_i; r') \; \textup{ for some}
\; r'\}| - |\{z_i | y_i=E^i(x_1, \ldots, x_i; z_i; r)\}| \geq
\Gamma.
\end{equation}
\\
The bars $|\cdot |$ indicate size of the enclosed set. For a
nonrandom stream cipher, the keystream $z_i$ is uniquely fixed by
the plaintext vector $\mathbf{x}_i$ and the ciphertext symbol
$y_i$. Therefore, if the randomizer in
(\ref{Gammarandomcipherdef}) is ignored so that it applies to a
nonrandom cipher, a nonrandom cipher would have $\Gamma = 0$. Note
that the sets whose sizes appear in the above equation, both for
random ciphers and their nonrandom reductions, are constructed
only on the basis of the $i$-th ciphertext symbol $y_i$, and not
on the basis of the entire ciphertext sequence. Thus, the
definition of $\Gamma$ only gives the \emph{number of possible
keys per symbol of ciphertext} under known-plaintext attack, while
the number of possible keys based on the entire ciphertext
sequence (that is illustrated schematically by the overlap sets in
Fig.~1) may be significantly less. In this sense, our definition
has a restricted symbol by symbol scope but is easy to calculate
with, similar to the independent particle approximation in
many-body physics. It does not by itself determine the precise
security of the cipher, but rather is the starting point of
precise analysis, which is a difficult task just as correlations
in interacting many-body systems are always difficult to deal with
in a rigorous quantitative manner.

It is possible to satisfy the random cipher condition
(\ref{random}) with $\Gamma =0$. This happens, e.g., when
(\ref{Gammarandomcipherdef}) holds for some ciphertext symbols
with $\Gamma >0$ but some others with $\Gamma=0$, so the overall
condition (\ref{Gammarandomcipherdef}) is only satisfied for
$\Gamma=0$. A different measure  of randomization $\Lambda$,
bearing directly on (\ref{random}), may be introduced which has
the property that $\Lambda=0$ is equivalent to a nonrandom cipher.
For the case where the ciphertext alphabet is finite and for given
$\mathbf{x}_i,z_i$ and $r$, let
\begin{equation} \label{lambdarandomcipherdef}
\Lambda=|\{ y_i | y_i = E^i(x_1, \cdots , x_i;z_i;r') \; \textup{ for some}
 \; r' \}| - |\{ y_i | y_i = E^i (x_1, \cdots , x_i;z_i;r)
\}|.
\end{equation}
Thus, condition (\ref{random}) is equivalent to $\Lambda >0$ for
some $\mathbf{x}_i,z_i$ and $r$. It follows that $\Lambda=0$ for
all $(\mathbf{x}_i,z_i)$ is equivalent to the cipher being
nonrandom. $\Lambda+1$ is the number of possible output signal
symbols corresponding to a given input symbol and running key
value. Thus, the parameter $\Lambda$ measures directly the degree
of per symbol ciphertext randomization, while $\Gamma$ measures
the per symbol key redundancy. It is possible that a $\Gamma=0$
random cipher is still useful due to the additional loads on Eve
to record and store more information from her observation. On the
other hand, for the \emph{typical} case where $z_i$ is in
one-to-one correspondence with $y_i$ for given $\mathbf{x}_i$ and
$r$, $\Gamma >0$ implies $\Lambda>0$ for every $\mathbf{x}_i$ and
$z_i$, which in turn implies that a cipher with $\Gamma
> 0$ is random in the sense of (\ref{random}). A simple
application of the $\Gamma$ and $\Lambda$ characterizations to
$\alpha\eta$ leads to information-theoretic lower bounds on the
unicity distances $n_0$ and $n_1$ for CTA and KPA, as discussed in
Sec. 4.3. The following simple example also serves to illustrate
the above definitions:
\newline\newline
\textbf{Example} (Random cipher) \newline Let
$\mathcal{X}=\{0,1\}$, $\mathcal{K}=\{k_0,k_1,k_2,k_3,k_4\}$ and
$\mathcal{Y}=\{a,b,c,d,e\}$. Fig.~2 lists the possible ciphertexts
for each plaintext and key pair.

\begin{figure*}[htbp]
\begin{tabular} {|c|c|c|} \hline  $x$ & $k$ & $y$\\
\hline \hline$0$ & $k_0$ & $a,b$\\
\hline $1$&$k_0$ & $c,d,e$\\
\hline$0$ & $k_1$ & $c,d$\\
\hline$1$ & $k_1$ & $e,a,b$\\
\hline$0$ & $k_2$ & $e,a$\\
\hline$1$ & $k_2$ & $b,c,d$\\
\hline$0$ & $k_3$ & $b,c$\\
\hline$1$ & $k_3$ & $d,e,a$\\
\hline$0$ & $k_4$ & $d,e$\\
\hline$1$ & $k_4$ & $a,b,c$\\ \hline
\end{tabular}
\caption{Encryption table for a simple random cipher.}
\end{figure*}
 For this cipher, one can easily verify that at
least 2 key values connect every possible plaintext-ciphertext
pair. In addition, every plaintext-key pair can lead to at least
two different ciphertexts. In terms of the definitions given
above, this cipher
has $\Gamma=1$ and $\Lambda=1$. 
\newline

\section{Quantum Random Ciphers}

The known and possible advantages of a random classical cipher
over a nonrandom one were discussed in the previous section. While
it is possible to implement a random cipher classically using
random numbers generated on Alice's side, this is not currently
practical at high ($\sim$ Gbps) rates. As will become clear in the
sequel, the quantum encryption protocol $\alpha\eta$ (Various
implementations are described in
\cite{barbosa03,corndorf03,pra05,ptl05,hirota05} - The protocol in
\cite{hirota05} is a variation on the original $\alpha\eta$ of
\cite{barbosa03}) effectively implements a random cipher from
Eve's point of view for a given choice of her measurement, the
difference from a classically random cipher being that it uses
coherent-state quantum noise to perform the needed randomization.
Before we describe $\alpha\eta$, we define some concepts that
capture the relevant features of a quantum random cipher. As
emphasized in Section 2.2, we will confine our attention to
\emph{stream} ciphers. First, we straightforwardly extend the
usual stream cipher to one where the ciphertext is a quantum
state. Our motivation for this definition is that, from the point
of view of the legitimate users Alice and Bob, $\alpha\eta$ is a
quantum stream cipher with negligible $\lambda$ in the sense given
below:
\\ \\
\textbf{Definition}\ \emph{($\lambda$-Quantum Stream Cipher (QSC)})\textbf{:}\\
A quantum stream cipher is a cipher for which the following two
conditions are satisfied:
\begin{enumerate} [A.]
\item{ The encryption map $e_k(\cdot)$ takes the $n$-symbol
plaintext sequence $\mathbf{x}_n$ to a quantum state $n$-sequence
$\mathbf{\rho}$ in the $n$-fold tensor product form:

\begin{equation} \label{quantcipher}
\mathbf{\rho}= e_k(\mathbf{x}_n) = \rho_{1}(x_1;z_1) \otimes
\ldots \otimes \rho_{n}(x_1, \ldots ,x_n;z_n),
\end{equation}
and}
\item{ Given the key $k$, there exists a measurement on the
encrypted state sequence, that recovers each  plaintext symbol
$x_i$ with probability $P_{dec} > 1 - \lambda$.}
\end{enumerate}

Here, as in Section 2.2, $(Z_1, \ldots, Z_n)$ is the keystream
generated from the seed key $K$.  A few comments will help clarify
the definition. First, note that the tensor product form of the
state in condition A retains for a quantum cipher the property of
a classical cipher that one can generate the components in the
$n$-sequence of states that constitute the output of a cipher one
after the other in a time sequence. Note also that, analogous to a
classical stream cipher, the $i$-th tensor component of $\rho$
depends on just $z_i$ and  $(x_1,\ldots,x_i)$. Condition B is the
generalized counterpart of the decryption condition
Eq.(\ref{decryption}) for a classical cipher -- we now allow a
small enough decryption error probability. Thus, the per-symbol
error probability is bounded above by $\lambda < 1$.

We now want to bring the concept of classical \emph{random} cipher
defined in the previous section into the quantum setting. Our
motivation in doing so is to show that, for an attacker making the
same measurement on a mode-by-mode basis without knowledge of the
key, $\alpha\eta$ reduces to an equivalent $\Gamma$-Random Cipher
with significantly large $\Gamma$. Since the output of a quantum
cipher is a quantum state and not a random variable, we will need
to specify a POVM $\{\Pi_{\mathbf{y}_n}\}$ whose measurement
result $\mathbf{Y}_n$ supplies the classical ciphertext.  Note
that in this quantum situation different choices of measurement
may result in radically different kinds of ciphertext. Note also
that the user's and the attacker's measurements may be different.
Our definition of a quantum random stream cipher below will apply
relative to a chosen ciphertext $\mathbf{Y}_n$ defined by its
associated POVM. We will also assume that, from the eavesdropper's
viewpoint, the same measurement is made on each of the $n$
components of the cipher output. In other words, the POVM defining the
ciphertext $\mathbf{Y}_n$ is a tensor product of identical POVMs $\{\pi_y\}$.
\\ \\
\textbf{Definition} (\emph{$(\Gamma,\lambda,\lambda',\{\pi_y\})$- Quantum Random Stream Cipher (QRC)})\textbf{:}\\
An $(\Gamma,\lambda,\lambda',\{\pi_y\})$ - quantum random stream
cipher is a $\lambda$-quantum stream cipher such that for the
ciphertext given by the result of the product POVM
$\{\Pi_{\mathbf{y}_n}= \bigotimes_{i=1}^{i=n} \pi_{y_i}\}$,
\begin{enumerate}[A.]
\item
one has an $\Gamma$-random stream cipher satisfying
Eq.(\ref{Gammarandomcipherdef}), and
\item
the probability of error per symbol $P_{dec}'$ using the key
\emph{after} measurement is $P_{dec}' > 1 -\lambda'$.
\end{enumerate}

Several comments are given to explain this definition:
\begin{enumerate}[1.]
\item{ While condition QRC-B above appears similar to the condition QSC-B for a
quantum stream cipher, there is a crucial difference. In the
latter, the decryption probability $P_{dec}$ takes into account
the possibility that the \emph{quantum measurement} (as well as
classical post-processing) made on the cipher state can depend on
the key, i.e. it refers to Bob's rather than Eve's error
probability. In QRC-B, we are considering the probability of error
involved for Eve when she decrypts using a quantum measurement
 independent of the key followed by classical post-processing that is , in general, ``collective'' and depends on the key. Thus,
the parameter $\lambda'$ is related to the symbol error
probability under this latter restriction while the parameter
$\lambda$ in QSC-B is tied to the symbol error probability for a
quantum measurement allowed to depend on the key. We see that
there are two measurements implicit in our definition of a QRC -
one made by the user with the help of the key, and the other given
by $\{\pi_y\}$ made by the attacker without the key. See also Item
3 below. As we shall see, $\alpha\eta$ satisfies QRC-B with
negligible $\lambda'$ under a heterodyne or phase measurement attack by
Eve.}
\item{ $\Gamma$ in QRC-A, as in Eq.(\ref{Gammarandomcipherdef}), is a measure of the
'degree of intermixing' of the regions of ciphertext space
corresponding to different key values on a symbol-by-symbol basis.
If $\{\pi_y\}$ describes a discrete measurement, a $\Lambda$
corrresponding to Eq.(\ref{lambdarandomcipherdef}) can also be
introduced.}
\item{ Our stipulation that the same POVM be measured on each of
the components of the cipher output is tantamount to restricting
the attacker to identical measurements on each tensor component
followed by collective processing. We will call such an attack a
\emph{collective attack} in this paper (also in \cite{yuen05qph}).
This definition is different from the usual collective attack in
quantum cryptography \cite{gisin02}: in the latter, following the
application of identical probes to each qubit/qumode, a joint
quantum measurement on all the probes is allowed. 
In our case, there is no probe for Eve to set as we conceptually
allow her a full copy of the quantum state. Doing so, we can upper
bound her performance. (This is an important feature of our
so-called KCQ approach to encryption and key generation. See
\cite{yuen03} for discussion.) Thus, allowing a joint
measurement, as also nonidentical measurements on each output
component, will be called a joint attack.}

\item{In analogy with the classical random cipher definition Eq.~(\ref{Gammarandomcipherdef}), one
may wonder why the private randomizers $R_i$ used in that
definition are missing from that of the quantum random cipher.
Indeed, one may randomize the quantum state $\rho_i(x_1, \ldots
,x_i;z_i)$ to $\rho_i(x_1, \ldots ,x_i;z_i;r_i)$ using a private
random variable with probability distribution $p_{r_i}$. However,
since the value of $R_i$ remains unknown to both user and attacker
(Indeed, the user should not need to know $R_i$ in order to
decrypt or even to encrypt in the case of $\alpha\eta$), one sees
that all probability distributions of Bob's or Eve's measurements
in this situation are given by the state $\rho'_i(x_1, \ldots
,x_i;z_i)=\sum_{r_i}{p_{r_i}\rho_i(x_1, \ldots ,x_i;z_i;r_i)}$, in
which there is no explicit dependence on $r_i$. In particular, we
mention here that exactly such quantum state randomization, called
Deliberate Signal Randomization (DSR), has been proposed in the
context of $\alpha\eta$ in \cite{yuen03} for the purposes of
enhancing the information-theoretic security of $\alpha\eta$.}
\item{It is important to observe that the definitions given above
both for classical and quantum random ciphers are not arbitrary
ones, but rather the mathematical characterizations of very
typical situations involving randomization in classical and
quantum cryptosystems.}
\end{enumerate}

We present an example of a QRC in the next section: the
$\alpha\eta$ cryptosystem.

\section{The $\alpha\eta$ cryptosystem}

\subsection{Operation}

We now describe the $\alpha\eta$ system and its operation as a
quantum cipher:
\begin{enumerate}[(1)]

\item
Alice and Bob share a secret key $\mathbf{K}_s$.

\item
Using a \emph{key expansion function} $ENC(\centerdot)$, e.g., a
linear feedback shift register or AES in stream cipher mode, the
seed key $\mathbf{K}_s$ is expanded into a running key sequence
that is chopped into $n$ blocks:
$\mathbf{K}_{Mn}=ENC(\mathbf{K}_s)=(K_1, \ldots , K_{mn})$. Here,
$m=\log_2(M)$, so that $Z_i \equiv (K_{(i-1)m +1}, \ldots,
K_{im})$ can take $M$ values. The $Z_i$ constitute the
\emph{keystream}.

\item
The encrypted state $e_{\mathbf{K}_s}(\mathbf{X}_n)$ of
Eq.(\ref{quantcipher})is defined as follows. For each bit $X_i$ of
the plaintext sequence $\mathbf{X}_n = (X_1, \ldots, X_n)$, Alice
transmits the \emph{coherent state}
\begin{equation} \label{state}
|\psi(X_i,Z_i)\rangle=|\alpha e^{i\theta(X_i,Z_i)}\rangle.
\end{equation}
Here, $\alpha \in \mathbb{R}$ and $\theta(X_i,Z_i)$ takes values
in the set $\{0,\pi/M,\ldots,(2M-1)\pi/M\}$. The function $\theta$
taking the data bit and keystream symbol to the actual angle on
the coherent state circle is called the \emph{mapper}. In this
paper, we choose $\theta(X_i,Z_i)=[Z_i/M+(X_i\oplus
Pol(Z_i))]\pi$. $Pol(Z_i)= 0$ or $1$ according to whether $Z_i$ is
even or odd. This distribution of possible states is shown in
Fig.~2. Thus $K_i$ can be thought of as choosing a `basis' with
the states representing bits $0$ and $1$ as its end points. In
general, one has the freedom to vary the mapper in various ways
for practical reasons. See, e.g, \cite{pra05}.

\item
In order to decrypt, Bob runs an identical ENC function on his
copy of the seed key. For each $i$, knowing $Z_i$, he makes a
quantum measurement to discriminate just the two states
$|\psi(0,Z_i)\rangle$ and $|\psi(1,Z_i)\rangle$.
\end{enumerate}

\begin{figure*} [htbp]
\begin{center}
\rotatebox{-90} {
\includegraphics[scale=0.7]{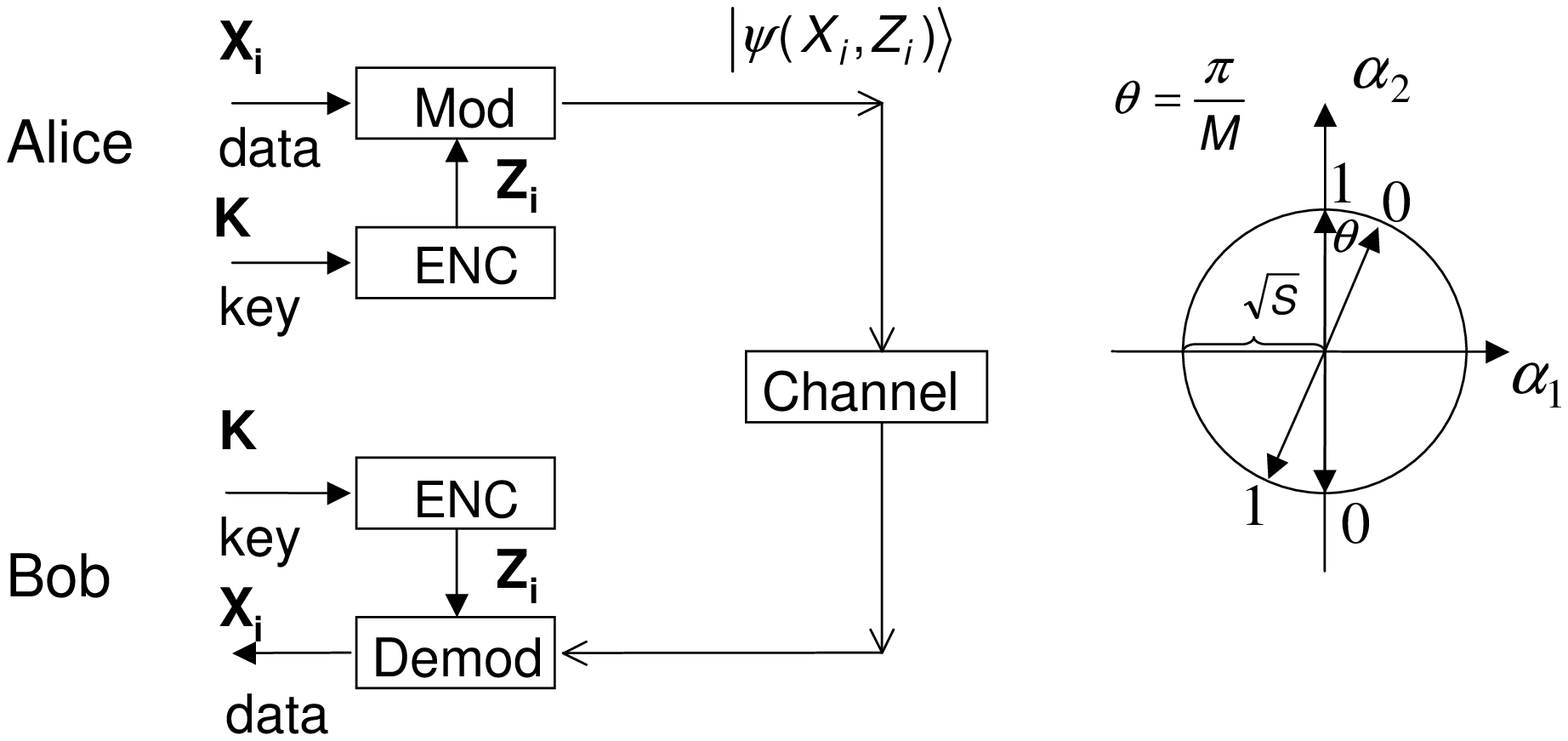}}
\caption{Left -- Overall schematic of the $\alpha\eta$ encryption
system.
 Right -- Depiction of two of $M$ bases with interleaved logical
bit mappings.}
\end{center}
\end{figure*}


To decrypt in step (4) above, Bob, in general would need a phase
reference. This is effectively provided by the use of Differential
Phase Shift Keyed (DPSK) signals in the implementations of
$\alpha\eta$. See \cite{corndorf03,pra05,ptl05} for details. Doing
so does not compromise security as we still assume that Eve has a
perfect copy of the transmitted state.

If the line transmittance between Alice and Bob is $\eta$, Bob
receives a coherent state with energy $\eta S$ instead of $S
\equiv |\alpha|^2$. The optimal quantum measurement
\cite{helstrom76} for Bob has error probability
\begin{equation} \label{pB}
P^B_e \sim \frac{1}{4} \exp(-4\eta S).
\end{equation}
It is thus apparent that $\alpha\eta$ is a $\lambda$-quantum
cipher in the sense of Section 3 with  $\lambda \sim \frac{1}{4}
\exp(-4\eta S)$. For the $S \sim 4 \times 10^4$ of \cite{pra05},
over a distance of 80 km at a loss of 0.2 dB/km, we have $\eta S
\sim 10^3$ photons. For this mesoscopic level, $\lambda$ is $\sim
\exp(-1000)$, which is completely negligible compared, say, to the
standard acceptable BER limit of $10^{-9}$, which arises from
device imperfections, for an uncoded optical on-off keyed line.

Let us briefly indicate how this system may provide data security
by considering an \emph{individual attack} on each data bit $X_i$
by Eve. Under such an attack, one only looks at the per-bit error
probability ignoring correlations between the bits. Under this
assumption, Eve, not knowing $Z_i$, is faced with the problem of
distinguishing the density operators $\rho^0$ and $\rho^1$ where
\begin{equation} \label{rho}
\rho^b=\sum_{Z_i}\frac{1}{M}|\psi(b,Z_i)\rangle\langle\psi(b,Z_i)|.
\end{equation}
For a fixed signal energy $S$, Eve's optimal error probability is
numerically seen to go asymptotically to $1/2$ as the number of
bases $M \rightarrow \infty$ (See Fig. 1 of \cite{barbosa03}). The
intuitive reason for this is that increasing $M$ more closely
interleaves the states on the circle representing bit 0 and bit 1,
making them less distinguishable. Therefore, at least under such
individual attacks on each component qumode \footnote{When
referring to an optical field mode, we use the term \emph{qumode}
(for 'quantum mode', in analogy to 'qubit').} of the cipher
output, $\alpha\eta$ offers any desired level of security
determined by the relative values of $S$ and $M$. While we are not
concerned in this paper with key generation, it may be observed
that unambiguous state determination (USD) attacks on $\alpha\eta$
are totally ineffective due to the large number of $2M$ states
involved.

In our security analysis, Eve is always assumed to be at the
transmitter so that $\eta=1$ for her. Without knowing the key,
however, her performance on the data is still poor as described in
the above paragraph. Her attacks on the key are described in the
following. We have assumed that the users can utilize the signal
energy $\eta S$ to maintain a proper bit error rate without
channel coding, despite possible interference from Eve. This does
not place a stringent requirement on $\eta$ itself as one can
typically go around 80 km in fiber before the signal needs to be
amplified. In case Eve's interference is too strong and causes
error, it would be detected in a message authentication code which
always goes with encryption. There is clearly no need to do
separate intrusion detection in this direct encryption case, but
it turns out there is also no need in the key generation regime
\cite{yuen05qph,yuen03} which we do not discuss in this paper.

\subsection{$\alpha\eta$ as a Random Cipher}
We showed in the previous subsection that $\alpha\eta$ may be
operated in a regime of $S$, $\eta$ and $M$ where it is a
$\lambda$-quantum cipher for $\lambda \sim 0$. We now show, that
from Eve's point of view, under both a heterodyne and phase
measurement attack, $\alpha\eta$ appears effectively as a quantum
\emph{random} cipher according to the characterization of Section
3. Note that the randomization in $\alpha\eta$ can also be
effected in principle by using an additional classical random
number generator. This is not required in $\alpha\eta$ as
high-speed randomization is automatically provided by the
coherent-state quantum noise.

To see the quantum random cipher characteristic of $\alpha\eta$,
consider employing the following two measurements for obtaining
$\{\pi_y\}$ in the quantum random cipher definition:
\begin{enumerate}[1)]
\item{ (Heterodyne measurement) $\pi_y = \frac{1}{\pi} |y\rangle\langle y|, y \in
\mathbb{C}.$}
\item{ (Canonical Phase measurement) $\pi_{\theta} = \frac {1} {2\pi} \sum_{n,n'=0}^{\infty} e^{i(n-n')\theta} |n\rangle \langle n'|, \theta \in [0,2\pi).$
}
\end{enumerate}
To show that the conditions for a QRC are satisfied, let us first
consider QRC-B. It may be  shown \cite{yuen03} that the error
probabilities $\lambda'$ involved are respectively $\sim
\frac{1}{2}e^{-S}$ and $\sim \frac{1}{2}e^{-2S}$ for the
heterodyne and phase measurements.

Turning to QRC-A, let us estimate the value of $\Gamma$ under
heterodyne and phase measurement. For a signal energy $S$, the
heterodyne measurement is Gaussian distributed around the
transmitted amplitude with a standard deviation of $1/2$ for each
quadrature while the phase measurement has an approximately
Lorentzian distribution around the transmitted phase with standard
deviation $\sim 1/{\sqrt{S}}$. If we assume that, given a certain
transmitted amplitude/phase, the possible ciphertext values are
uniformly distributed within a standard deviation on either side
and ciphertext values outside this range are not reached (this
will be called the \emph{wedge approximation}), we get the
following estimates $N_{het}$ and $N_{phase}$ for the number of
keystream values $z_i$ covered by the quantum noise under
heterodyne and phase measurements:
\begin{equation} \label{numberestimate}
N_{het}=2N_{phase} = M/(\pi \sqrt{S}).
\end{equation}
If the value of the randomizer $R$ is fixed (corresponding to
rotation by a given angle within the wedge), $Z_i$ is fixed by the
plaintext and ciphertext. Thus we have according to
Eq.~(\ref{Gammarandomcipherdef}) that
\begin{equation}\label{Gammahet}
\Gamma_{het}=N_{het} -1 \cong M/(\pi \sqrt{S}),
\end{equation}
and that
\begin{equation} \label{Gammaphase}
\Gamma_{phase} \cong \Gamma_{het}/2 \cong M/(2 \pi \sqrt{S}).
\end{equation} As expected, the $\Gamma$'s of both measurements
increase as the number of bases $M$ increases, and decrease with
increasing signal energy $S$ that corresponds to decreasing
quantum noise. For example, using the experimental parameters in
\cite{pra05} of $S \sim 4 \times 10^4$ photons and $M \sim 2
\times 10^3$ has $\Gamma_{het} \sim 3$. The $\Lambda$ (cf.
Eq.~(\ref{lambdarandomcipherdef}) characteristics of $\alpha\eta$
will be considered in Sec.~5.2 in connection with the Nishioka
group attack. The relevance of these parameters for security is
considered in detail in the next subsection and in Sec.~5.2.

\subsection{$\alpha\eta$: Information-theoretic and Complexity-Theoretic Security}
Before discussing $\alpha\eta$ security, we comment that
$\alpha\eta$ direct encryption is often compared to BB84 key
generation followed by the use of the generated key in either
one-time pad or a standard cipher like AES. This is not an
appropriate comparison because $\alpha\eta$ already assumes that
the users share a key. Perhaps the source of the confusion is that
both $\alpha\eta$ and BB84 involve the use of quantum states. In
any case, the appropriate comparison would be between $\alpha\eta$
and a standard cipher like one-time pad or AES - we do make such a
comparison in the following.

We will consider in turn the information-theoretic (IT) and
\emph{complexity-theoretic (CT)} security of $\alpha\eta$. In
standard cryptography, no rigorous result is known about the
quantitative security level of any cipher, save the one-time pad.
Since $\alpha\eta$ includes a classical stream cipher ENC (See
Fig. 1), we may in general expect a similarly murky state of
affairs regarding its quantitative security. However, it will turn
out that, under known-plaintext attacks, one can claim
\emph{additional} security from the physical coherent-state noise
for a suitably modified $\alpha\eta$ with any cipher ENC, as
compared to ENC alone.

\subsubsection{Information-theoretic (IT) Security: Qualitative discussion}

Considering first IT security, we discuss in turn qualitatively
the cases of ciphertext-only, known-plaintext, and statistical
attacks on the data as well as the key. Subsequently, for the
former two cases, we give lower bounds for the unicity distances
$n_0$ and $n_1$ (See Appendix A for definitions).

As mentioned in Appendix A, for a nondegenerate ENC box cipher,
one can protect the key completely and attain data security up to
the Shannon limit under CTA.  If the same ENC box is used in
$\alpha\eta$ one may consider, as in Sec. 4.1, an attack in which
Eve attacks each data bit using only the measurement result from
the corresponding qumode. Although under such an assumption IT
security obtains as $M/\sqrt{S} \rightarrow \infty$, this attack
is too restrictive since Eve does gain information on the key from
each qumode measurement that could be useful in learning about
other data bits as well. Such attacks utilizing key correlations
across data bits may be launched against standard stream ciphers.
Under the wedge approximation of Sec.~4.2, Eve is able to narrow
her choice of basis down to $\Gamma$ possible values. Even if
$\Gamma$ is large, the key security (and hence data security) is
not as good as that of the ENC box alone for which case the
keystream bits are \emph{completely} random to Eve. However, one
can still derive a unicity distance lower bound (See below). This
defect of $\alpha\eta$ may be removed by the use of Deliberate
Signal Randomization (DSR) introduced in \cite{yuen03}. However,
the concrete analysis of systems using various forms of DSR are
still under progress. But see \cite{yuen06pla}.

Let us now consider the case of known-plaintext attacks on the
key. As discussed in Appendix A, most nonrandom ciphers have a
nondegeneracy distance $n_d$ at which the key is fixed under a
known-plaintext attack. We also mentioned that for random ciphers,
such a distance may not exist, so that it is unknown whether or
not they possess IT security against KPAs. Since $\alpha\eta$ is
random, the same remark applies to it. However, a finite unicity
distance $n_1$ may exist for $\alpha\eta$ and other random ciphers
beyond which the key is fixed in a KPA. While rigorous analysis is
difficult and is so far limited to the unicity distance bound
given below, we believe that such is the case for the original
$\alpha\eta$ with no modification, so that it has no IT security
for large enough $n$.

The statistical attacks fall between the above two extremes. Thus,
there may exist a crossover point where $\alpha\eta$ security
becomes better than that of the ENC box alone as one moves from
CTA towards KPA. However, no quantitative results, e.g., the
unicity distance under STA, are known.  To summarize, we believe
that under all cryptographic attacks, $\alpha\eta$ has no IT
security for large enough $n$, i.e., $\lim _{n \rightarrow \infty}
H(K|\mathbf{Y}_n^E) = 0$. However, the use of $\alpha\eta$ should
extend the unicity distance beyond that of the cipher ENC used in
it for some statistical attacks and for known-plaintext attacks.

\subsubsection{Information-Theoretic (IT) Security: Unicity Distance Lower Bounds}

Nonrigorous estimates of the unicity distance $n_1$ against KPA
for standard stream ciphers are often made via a capacity argument
in the so-called ``correlation attacks'' (See, e.g.,
\cite{chepyzhov00}). The bound
\begin{equation}
\label{capacitybound} n \geq |K|/C,
\end{equation}
where $C$ is the capacity of Eve's effective channel, follows from
the converse to the coding theorem \cite{cover91}. The application
of (\ref{capacitybound}) to correlation attacks is nonrigorous
because the assumption of independent noise in each bit is not
valid. In the case of $\alpha\eta$, \emph{rigorous} lower bounds
on $n_0$ and $n_1$ can be obtained from (\ref{capacitybound})
because of the independent qumode to qumode coherent-state noise.
Under the wedge approximation to the noise distribution for
evaluating Eve's capacity in (\ref{capacitybound}), it may be
shown \cite{eguchi06} that for uniform data, the CTA unicity
distance
\begin{equation} \label{n0bound}
n_0 \geq \frac{|K|}{\log_2(\frac{M}{\Lambda +1})},
\end{equation}
and for KPA,
\begin{equation} \label{n1bound}
n_1 \geq \frac{|K|}{\log_2(\frac{M}{\Gamma +1})}.
\end{equation}\newline
In terms of the experimental parameters of \cite{pra05}, this
gives $n_0 \geq 550, n_1 \geq 490$. While these are much bigger
than $n_0 \sim 120$ bits for English, no precise practical
conclusion can be drawn, both because they are just lower bounds
and because the actual complexity of key determination as a
function of $n$ is not yet known. For the numbers above, the
cryptosystem would be secure if the optimal complexity is
exponential in $n$.

\subsubsection{Complexity-theoretic (CT) Security}

Apart from IT security, the issue of complexity-theoretic (CT)
security is of great practical importance. Indeed, in
\cite{yuen05qph}, we have argued that large enough search
complexity security is as good as information-theoretic security
in reality. For standard ciphers, we have seen that there is no IT
security beyond the nondegeneracy distance. Thus, standard ciphers
rely for their security under KPA basically on the complexity of
algorithms to find the key. We now compare the situation with that
of $\alpha\eta$. For any attack, the mere fact that
$H(K|\mathbf{Y}_n^E)=0$ (for CTA and STA) or $H(K|\mathbf{Y}_n^E
\mathbf{X}_n)=0$ (for KPA) does not mean that the unique key can
be readily obtained from $\mathbf{Y}_n^E$ (and $\mathbf{X}_n$ in
the case of KPA). For most ciphers, one needs to run an algorithm
to obtain it. At worst, this algorithm can be a \emph{brute force
search} - one decrypts $\mathbf{Y}_n^E$ with all the $2^{|K|}$
possible keys until a valid plaintext is obtained. This search can
easily be made prohibitive by choosing $|K|$ large enough --
$|K|\sim 4000$ used in experimental $\alpha\eta$ \cite{pra05} is
already way beyond conceivable search capability. A better
procedure that we call an \emph{assisted brute force search} can
exploit partial knowledge of the possible running key values for
each bit as follows. Since each basis is specified by
$m=\log_2(M)$ bits of the running key, and the seed key is
revealed by a $|K|$-bit sequence of the running key for an ENC box
of Fig.~3 that is an LFSR with known connection polynomial, we
obtain an \emph{assisted brute-force search complexity} of
\begin{equation}\label{searchcomplexity}
\mathcal{C}=\Gamma^{|K|/m}.
\end{equation}
For $|K|=4400$ used in \cite{pra05}, $\mathcal{C} \sim 2^{630}$
which is far beyond any conceivable search capability.  While it
is not known what Eve's \emph{optimal} search complexity is, the
advantage here is that this degree of randomization is achieved
automatically by the coherent-state quantum noise at the $\sim$
Gbps rate of operation of the system. Note also that it is not
hard to increase $M$ while maintaining the same data rate because
the number of bits needed to select a basis on the circle scales
logarithmically with $M$.

In practice, heuristic algorithms based on the structure of the
ENC cipher are used to speed up the search. The rigorous
quantitative performance of these algorithms is unknown for
standard ciphers. However, one may view $\alpha\eta$ as an
``enhancer'' of security by providing an additional `physical
encryption' on top of the standard `mathematical encryption'
provided by the ENC box as follows.

For the ENC of Fig.~3 used as a standard cipher, so that
\begin{equation} \label{ENC}
Y_i=X_i\oplus K_i, \ K_i=ENC(\mathbf{K}_s),
\end{equation}
let the unicity distance for KPA be $n_1$ . Let us assume that
there exists an algorithm ALG($Y_{n_1},X_{n_1})$) whose output is
the seed key $K_s$ and that ALG has complexity $C$ when used with
inputs of length $n_1$. In order to compare this complexity with
that of $\alpha\eta$, we assume that the same ENC is used in an
$\alpha\eta$ system. However, since $m$ bits of the keystream
output of ENC are used to choose the basis for one data bit in
$\alpha\eta$, we first 'match' the data stream and keystream in
$\alpha\eta$ as follows.

We \emph{expand} the ENC output keystream by applying $m$
deterministic $m$-bit to $m$-bit functions $\{f_j\}_{j=1}^m$ to
each keystream symbol $Z_i$ to get a new keystream $\mathbf{Z'}$
as follows:
\begin{equation} \label{kprimeprime} Z' = (f_1(Z_1), \cdots,
f_m(Z_1), f_1(Z_2), \cdots, f_m(Z_2), \cdots).
\end{equation}
We then use  $Z'$ instead of $Z$ to choose the basis for each data
bit.

The above modification results in the $i$-th $m$-block of
ciphertext $Y_{(i-1)m} \cdots Y_{im}$ being dependent only on
$K_{(i-1)m} \cdots K_{im}$ and $X_{(i-1)m} \cdots X_{im}$ for
\emph{both} ENC and $\alpha\eta$ with ENC. Under a KPA on ENC
alone, using a known plaintext of length $n_1$, $K_1\ldots
K_{n_1}$ is known \emph{exactly}. For ENC augmented with
$\alpha\eta$ in the described manner, it \emph{may} happen that
because of the randomization of $Z'_1 \cdots Z'_{n_1}$, $K_1\ldots
K_{n_1}$ is not fixed by $\mathbf{Y}_{n_1}$ and
$\mathbf{X}_{n_1}$. In the latter case, we have IT security above
that of ENC alone, even though such security may be lost for large
enough $n$, as mentioned in the previous subsection.

Let us assume that, at the nondegeneracy distance $n_1$ of ENC,
$\alpha\eta$ with ENC does \emph{not} have IT security, so that
$H(K|\mathbf{X}_{n_1}\mathbf{Y}_{n_1})=0$. Assume also that
$n_1=mk$. Even in such a case, it appears harder to implement the
algorithm ALG that finds the key. As discussed in Section 2.2, the
reason is that the randomization of the ciphertext $Y_i$, for each
$i$, leaves each $Z_i$ undetermined immediately after the
measurement, even though, by our present assumption, only one
possible seed key $K$ can lead to the observed measurement
results. If the number of possibilities for each $Z_i$ is $l$, Eve
may need to run the algorithm ALG $l^{k}$ times resulting in a
complexity of $l^{n_1/m}C$ versus $C$ for ENC alone. Of course,
there may exist a clever algorithm that enables her to do much better. All
we claim here is that $\alpha\eta$ provides an additional but
unquantified layer of security over that of the ENC box against
KPA, both in the IT and CT senses. Thus, $\alpha\eta$ can be run
on top of any standard cipher in use at present, e.g. AES
(Advanced Encryption Standard), and provides an additional,
qualitatively different layer of physical encryption security over
AES under a known-plaintext attack.

An interesting point is that, if the above level of CT security
against known-plaintext attack is sufficiently high for some data
length $n$, there is at least as much security against CTA for the
same $n$. However, this comparison may not be practically
meaningful as a CTA can typically be launched for the entire
sequence of data while usually only a much smaller segment of
known-plaintext is available to the attacker. Typically, this
would imply the attacks can be parallelized. On the other hand,
the situation is practically favorable with AES used in the ENC -
see ref. \cite{yuen06pla}, where the immunity of $\alpha\eta$
against fast correlation attacks with and without DSR are also
treated.

\subsection{Overview of $\alpha\eta$ Features}
We summarize the main known advantages and rigorous security
claims regarding $\alpha\eta$ compared to previous ciphers:
\begin{enumerate} [(1)]
\item For known-plaintext attacks on the
key, $\alpha\eta$ using an LFSR has an additional brute force
search complexity given by $\Gamma^{|K|/m}$. When reconfigured as
in Sec. 4.3.3, it also has at least as much IT security as the ENC
box alone for the same length $n$ of data.

\item It may, when supplemented with further techniques \cite{yuen03}, have
information-theoretic security against known-plaintext attacks
that is not possible with nonrandom ciphers, and would also have
maximal information-theoretic security against ciphertext-only
attacks.

\item With added Deliberate Signal Randomization (DSR) \cite{yuen03}, it is expected to have improved information-theoretic
security on the data far exceeding the Shannon limit.

\item It has high-speed private true randomization (from quantum noise that even Alice does not know), which is not
possible otherwise with current or foreseeable technology.

\item It suffers no reduction in data rate compared to other known
random ciphers, because Bob needs to resolve only two and not $M$
possibilities (i.e, one data bit is transmitted per qumode).

\item It provides physical encryption, different from usual
mathematical encryption, that forces the attacker to attack the
optical line rather than simply the electronic bit output.
\end{enumerate}
\section{Nishioka et al's criticisms of $\alpha\eta$}

In this section, we discuss the criticisms made by Nishioka et al
\cite{nishioka04,nishioka05} and respond to them. This section has
some overlap with \cite{nair05} (that was not published), but
contains new material.

\subsection{Claims in Nishioka \emph{et al} \cite{nishioka05}}

Nishioka \emph{et al} claim that $\alpha\eta$ can be reduced to a
classical non-random stream cipher under the attack that we now
review. For each transmission $i$, Eve makes a heterodyne
measurement on the state and collapses the outcomes to one of $2M$
possible values. Thus, the outcome $j \in \{0, \cdots, 2M-1\}$ is
obtained if the heterodyne result falls in the wedge for which the
phase $\theta \in [\theta_j-\pi/2M, \theta_j+\pi/2M]$, where
$\theta_j= \pi j/M$. Further, for $q \in \{0, \cdots, M-1\}$
representing the $M$ possible values of each $Z_i$, Nishioka
\emph{et al} construct a function $F_j(q)$ with the property that,
for each $i$, and the corresponding running key value $Z_i$
actually used,
\begin{equation} \label{nishiokadecryption}
F_{j^{(i)}}(Z_i)=r_i
\end{equation}
with probability very close to 1. In fact, for the parameters
$S=100$ and $M=200$, they calculate the probability that
Eq.(\ref{decryption}) fails to hold to be $10^{-44}$, which value
they demonstrate to be negligible for any practical purpose.

The authors of \cite{nishioka05} further claim that the above
function $F_{j^{(i)}}(q)$ can always be represented as the XOR of
two bit functions $G_{j^{(i)}}(q)$ and $l_{j^{(i)}}$, where
$l_{j^{(i)}}$ depends \emph{only} on the measurement result. Thus,
they make the claim that the equation
\begin{equation} \label{reduction}
l_{j^{(i)}}=r_i \oplus G_{j^{(i)}}(Z_i)
\end{equation}
holds with probability effectively equal to 1. They then observe
that a classical additive stream cipher \cite{stinson} (which is
non-random by definition) satisfies
\begin{equation}\label{streamcipher}
l_i=r_i \oplus \tilde{k_i},
\end{equation}
where $r_i$, $l_i$, and $\tilde{k_i}$ are respectively the $i$th
plaintext bit, ciphertext bit and running key bit. Here,
$\tilde{k_i}$ is obtained by using a seed key in a
pseudo-random-number generator to generate a longer running key.
The authors of \cite{nishioka05} then argue that since
$l_{j^{(i)}}$ in Eq.(\ref{reduction}), like the $l_i$ in
Eq.(\ref{streamcipher}), depends just on the measurement result,
the validity of Eq.(\ref{reduction}) proves that the security of
Y-00 is equivalent to that of a classical stream cipher. In
particular, they claim that by interpreting $l_{j^{(i)}}$ as the
ciphertext, Y-00 is not a random cipher, i.e., it does not satisfy
Eq.(\ref{random}) of the next section.

We analyze and respond to these claims and other statements in
\cite{nishioka05} in the following section.

\subsection{Reply to claims in \cite{nishioka05}}

To begin with, we believe that Eq.~(\ref{decryption}) (Eq.~(14) in
\cite{nishioka05}) is correct with the probability given by them.
This content of this equation is simply that Eve is able to
decrypt the transmitted bit from her measurement data $J_N$ and
the key $\mathbf{K}_s$. In other words, it merely asserts that
Eq.(\ref{decryption}) holds for $\mathbf{Y}_N=J_N$. As such, it
does not contradict, and is even \emph{necessary}, for the claim
that $\alpha\eta$ is a random cipher for Eve. In fact, we already
claimed in \cite{yuen03} and \cite{yuen05pla} that such a
condition holds. In this regard, note also that the statement in
Section 4.1 of \cite{nishioka05} that ``informational secure key
generation is impossible when ( Eq.(\ref{decryption}) of this
paper) holds'' is irrelevant, since direct encryption rather than
key generation is being considered here. Furthermore, we have
already pointed out \cite{yuen05qph,yuen03,yuen05pla} that the
Shannon limit prevents key generation with the experimental
parameters used so far, a point missed in
\cite{nishioka04,nishioka05,loko05}. See also \cite{note1}.

We also agree with the claim of Nishioka \emph{et al} that it is
possible to find functions $l_{j^{(i)}}$ and $G_{j^{(i)}}(q)$, the
former depending only of the measurement result $j^{(i)}$, such
that Eq.(\ref{reduction}) holds, again with probability
effectively equal to one. The \emph{error} in \cite{nishioka05} is
to use this equation to claim, in analogy with
Eq.~(\ref{streamcipher}), that $\alpha\eta$ is reducible to a
classical nonrandom stream cipher.

To understand the error in their argument, note that, for
Eq.~(\ref{streamcipher}) to represent an additive stream cipher,
the $l_i$ in that equation should be a function \emph{only} of the
measurement result, and $\tilde{k_i}$ should be a function
\emph{only} of the running key. While the former requirement is
true also for the $l_{j^{(i)}}$ in Eq.~(\ref{reduction}), the
latter is certainly \emph{false} for the function
$G_{j^{(i)}}(Z_i)$ in Eq.~(\ref{reduction}), since it depends
\emph{both} on the measurement result $j^{(i)}$ and the running
key $Z_i$. Indeed, it can be seen that the definition of the
function $F_{j^{(i)}}(Z_i)$, and thus, $G_{j^{(i)}}(q)$ depends on
the sets $C_{j^{(i)}}^+$ and $C_{j^{(i)}}^-$ defined in Eq.~(12)
of \cite{nishioka05}. The identity of these sets in turn depends
on the relative angle between the basis $q$ and Eve's estimated
basis $\tilde{j^{(i)}}= j^{(i)} \bmod M.$ Thus, it is clearly the
case that $G_{j^{(i)}}(Z_i)$ must depend both on $j^{(i)}$ and
$Z_i$, a fact also revealed by the inclusion of the subscript
$j^{(i)}$ by the authors of \cite{nishioka05} in the notation for
$G$.

Notwithstanding the failure of Eq.~(\ref{reduction}) to conform to
the requirements of a stream cipher representation
Eq.~(\ref{streamcipher}), Nishioka \emph{et al} reiterate that
Y-00 is nonrandom because
\begin{equation} \label{l}
H(L_N|R_N, \mathbf{K}_s) =0
\end{equation}
holds, where $\mathbf{L}_N=(l_{j^{(1)}}, \ldots, l_{j^{(N)}})$.
This equation follows from Eq.~(\ref{reduction}) and so by
considering $\mathbf{L}_N\equiv \mathbf{Y}_N$ to be the
ciphertext, the Eq.(\ref{random}) is not satisfied, thus
supposedly making Y-00 nonrandom. The choice of $\mathbf{L}_N$ as
the ciphertext is supported by the statement in \cite{nishioka05}
that ``It is a matter of preference what we should refer to as
``ciphertext''.'' This is indeed true, especially considering that
there are different possible quantum measurements that may be made
on the quantum state in Eve's possession, each giving rise to a
different ciphertext. This point is also highlighted by our
definition of a qauntum random cipher.  However, if one wants to
claim equivalence to a non-random cipher for some particular
choice of ciphertext $\mathbf{Y}_N$, one must show that Eq.~(10)
is violated \emph{and} that Eq.~(11) is satisfied using the chosen
ciphertext in \emph{both} equations. In other words, no
equivalence to any kind of cipher is shown unless one can also
decrypt \emph{with the chosen ciphertext} and key alone. However,
one may readily see that, taking $\mathbf{Y}_N=\mathbf{L}_N$,
Eq.~(\ref{decrypt}) is not satisfied, i.e.,
 $H(\mathbf{R}_N|\mathbf{L}_N,\mathbf{K}_s)\neq 0$.
The reason is that, as we noted from our analysis above of the
function $G_{j^{(i)}}(q)$, decrypting $r_i$ requires knowledge of
certain ranges in which the angle between the basis chosen by the
running key and the estimated basis $\tilde{j^{(i)}}$ falls. To
convey this information \emph{for every possible} $j^{(i)}$, one
needs at least $\log_2(2M)$ bits. It follows that the single bit
$l_{j^{(i)}}$ is insufficient for the purpose of decryption, and
so Eq.~(\ref{decrypt}) cannot be satisfied for
$\mathbf{Y}_N=\mathbf{L}_N$. Therefore, we conclude, that in the
interpretation of $\mathbf{L}_N$ as the ciphertext, decryption is
not possible even if Eve has the key $\mathbf{K}_s$. Indeed, it is
$\mathbf{J}_N$ that can be regarded as a possible ciphertext,
since Eq.~(\ref{decrypt}) is satisfied for
$\mathbf{Y}_N=\mathbf{J}_N$. However, with this choice of
ciphertext, Y-00 necessarily becomes a \emph{random} cipher,
because $H(\mathbf{J}_N|\mathbf{R}_N,\mathbf{K}_s) \neq 0$, a fact
admitted by Nishioka \emph{et al} in \cite{nishioka05}.

We hope that the discussion above makes it clear that the
`reduction' of $\alpha\eta$ in \cite{nishioka05} to a non-random
cipher is false, and that in fact, no such reduction can be made
under the heterodyne attack considered in \cite{nishioka05}.
Indeed, as detailed in previous sections, the representation of
ciphertext by $\mathbf{Y}_N=\mathbf{J}_N$ does reduce it to a
\emph{random} cipher under the heterodyne attack. Its quantitative
random cipher characteristics, namely $\Gamma$ of
Eq.~(\ref{Gammarandomcipherdef}) and $\Lambda$ of
Eq.~(\ref{lambdarandomcipherdef}), are as follows, for various
definitions of ``ciphertext'' adopted.

If the full continuous observation on the circle is taken as the
ciphertext, then (\ref{Gammahet}) shows that $\Gamma \sim 3$ for
typical experimental parameters. If the ciphertext alphabet is
digitized and taken to be the $2M$ arc segments around the $2M$
states on the circle, then $\alpha\eta$ has, for any
$(\mathbf{x}_i,z_i,r)$, $\Lambda+1 =2(\Gamma+1)$ where $\Gamma$ is
given by (\ref{Gammahet}). If one attempts to `de-randomize' the
ciphertext by clubbing together the possibilities, $\Gamma$ would
increase while $\Lambda$ would decrease. In the nonrandom limit
where a fixed half-circle observation is taken to represent each
bit value, which is the nonrandom reduction discussed in
\cite{yuen05pla}, $\Gamma$ would increase from that of
Eq.~(\ref{Gammahet}) to $M$, making attacks on the key completely
impossible. On the other hand, while $\Lambda =0$ for a binary
ciphertext alphabet, the $2M$-outcome ciphertext would lead, from
Eq.~(\ref{Gammahet}), to an error probability per ciphertext bit
for Eve \cite{yuen05pla}:
\begin{equation}\label{wedgeerror}
P_b^E \sim 2/{\pi \sqrt{S}}. \end{equation} Eq.~(\ref{wedgeerror})
is obtained in the wedge approximation on a per qumode basis for
Eve, under the assumption that the state is uniformly distributed
on the circle which is satisfied for uniform data and an LFSR for
the ENC box of Fig.~3. It leads to $0.1-1\%$ error rate for Eve on
the ciphertext (not data \cite{note2}) for the experimental
parameters of \cite{barbosa03,pra05}. As a consequence, the data
security will far exceed the Shannon limit (\ref{shannonlimit})
because she would make many errors even when the correct key is
given to her for decryption. For any other ciphertext alphabet
division of the circle, it is clear that $\Lambda >0$ for any
$z_i$ and $\mathbf{x}_n$ from the same randomization for states
near the ciphertext alphabet boundaries on the circle.

In sum, there can be no nonrandom reduction of $\alpha\eta$. If
the ciphertext alphabet is chosen to make $\alpha\eta$ nonrandom,
then known-plaintext attack on the key is impossible and the
ciphertext itself would be obtained with significant noise.

We conclude this section by responding to some other statements
made in \cite{nishioka05}.

In Section 3.3, Nishioka \emph{et al} claim that ``The value of
$l_{j^{(i)}}$ does not have to be the same as that of
$l_{j^{(i')}}$ when $i \neq i'$, even if $j^{(i)}=j^{(i')}$
holds.'' This statement is in direct contradiction to their
previous statement in the same subsection that ``$l_{j^{(i)}}$
depends only on the measurement value $j^{(i)}$''.

In the same subsection, Nishioka \emph{et al} claim that ``In
(\cite{nishioka04}), we showed another concrete construction of
$l_{j^{(i)}}$ ...''. We could find no explicit construction of
$l_{j^{(i)}}$ in that paper. We were led to the choice of $l_i$
described in \cite{yuen05pla} by the attempt to make the stream
cipher representation Eq.~(\ref{streamcipher}) valid. In fact,
such a representation is claimed by Nishioka \emph{et al} in their
Case 2 of \cite{nishioka04}. It turned out, however, that
decryption using that $l_i$ suffered a $0.1-1$\% error depending
on the value of $S$ used as noted above. See \cite{yuen05pla} for
further details. While it was later claimed that they have a
different reduction in mind \cite{nishioka05}, the reduction in
\cite{yuen05pla} is the only one that makes $\alpha\eta$ nonrandom
(but in noise). In any case, as we have shown above, no
construction of a single-bit from the heterodyne or phase
measurement results can satisfy Eq.(\ref{decryption}) with the
extremely low probability given in \cite{nishioka05}.

\section{Acknowledgements}
We would like to thank Greg Kanter, Chuang Liang, and Koichi
Yamazaki for useful discussions. This work was supported by DARPA
under grant F30602-01-2-0528 and by AFOSR under grant
FA9550-06-1-0452.

\section*{Appendix A -- Security under Statistical and Known-Plaintext Attacks}

In this appendix, we summarize some relevant terminology and
results from ref. \cite{yuen05qph} on the key security of a random
cipher. We first present an overview of the various possible
cryptographic attacks possible on a cipher and some early results
on the subject. We also present our result on the security of a
nonrandom cipher under known-plaintext attacks. In the process, we
 define the important term `unicity distance' coined by
Shannon and broaden it to include the notion of `unicity distance
under known-plaintext attack' for both random and nonrandom
ciphers. We also define the important concept of `nondegeneracy'
for both random and nonrandom ciphers that is needed to make the
concept of unicity distance meaningful. Finally, we discuss how
random ciphers may enhance security against known-plaintext
attacks.

The following terminology in regard to cryptographic attacks has
bee used in this paper, as in \cite{yuen05qph}. This terminology
is not standard, however. In the cryptography literature, what we
call statistical attacks are sometimes referred to as
ciphertext-only attacks (See, e.g., \cite{stinson}, Ch. 2) but are
also often lumped together with known-plaintext attacks.

By a \emph{ciphertext-only attack (CTA)}, we refer to the case
where the probability distribution $p(\mathbf{X}_n)$ is completely
uniform, i.e., $p(\mathbf{X}_n) =2^{-n}$ to Eve, so that her
attack cannot exploit input frequencies or correlations and must
be based only on the ciphertext in her possession. By a
\emph{statistical attack (STA)}, we refer to the case where the
probability distribution $p(\mathbf{X}_n)$ is nonuniform, so that
Eve may in principle exploit input frequencies or correlations to
launch a better attack. Such an attack is typical when the
plaintext is in a language such as English. It is also the attack
that obtains when the $\{X_i\}$ are independent and identically
distributed (i.i.d.) but each $p(X_i)$ is nonuniform. By a
\emph{known-plaintext attack (KPA) } we mean the case where Eve
knows \emph{exactly} some length $m$ of plaintext $\mathbf{x}_m$.
Finally, by a \emph{chosen-plaintext attack (CPA)}, we mean a KPA
where the data $\mathbf{x}_m$ is chosen by Eve.

In standard cryptography, one typically does not worry about
ciphertext-only attack on nonrandom ciphers. The reason is that,
under CTA, Eq.~(\ref{shannonlimit}) is satisfied with equality for
large $n$ for the designed key length $ |K|=H(K)$ under a certain
`nondegeneracy' condition \cite{jkm} that is readily satisfied.
Thus, in practice, the data security is assumed to be sufficient
if $H(K)$ is chosen large enough by adjusting the key length. In
this paper, we would essentially make the same assumption and,
with few exceptions, do not discuss data security per se. However,
it follows from (\ref{shannonlimit}) that no meaningful lower
bound on $H(\mathbf{X}_n|\mathbf{Y}_n)$ exists for $n \gg |K|$. A
new fundamental treatment of data security in symmetric-key
ciphers has to be developed separately. Under CTA, it is also the
case for nonrandom nondegenerate ciphers that \cite{jkm}
\begin{equation}\label{keysecurity}
 H(K|\mathbf{Y}_n)=H(K),
\end{equation}
 i.e., the key is
\emph{statistically independent} of the ciphertext. Thus, no
attack better than pure guessing can be launched on the key.

The above two results do not hold for statistical and
known-plaintext attacks. Eve can indeed launch an attack on the
key and use her resulting information on the key to get at future
and past data. In fact, it is such attacks that are the focus of
concern for standard ciphers such as the Advanced Encryption
Standard (AES). For STAs, Shannon \cite{shannon49} characterized
the security by the so-called unicity distance. The \emph{unicity
distance}  $n_0$ of a cipher is the smallest input data length for
which $H(K|\mathbf{Y}_{n_{0}}) = 0$. In other words, if a
plaintext sequence of length $n_0$ is encrypted by the cipher, the
ciphertext contains enough information to fix the key (and hence,
the plaintext) uniquely -- the cipher has no information-theoretic
security. For nonrandom ciphers defined by Eq. (\ref{nonrandom}),
Shannon, in \cite{shannon49}, derived in terms of the data entropy
an estimate on $n_0$  that is independent of the cipher. This
estimate is actually \emph{not} a rigorous bound. Indeed, it can
be shown that one of the inequalities used in the derivation goes
in the wrong direction. Even so, the estimate works well
empirically for English language plaintexts, for which $n_0 \sim
25$ characters are found to be sufficient to break many ciphers.

We now consider, in some detail, security against known-plaintext
attacks. Here, a natural quantity to consider is
$H(K|\mathbf{X}_n\mathbf{Y}_n)$, since it provides a measure of
key uncertainty when both plaintext and ciphertext are known to
the attacker. Before we state the main result, we define the
notion of nondegeneracy distance. The reader can readily convince
himself that a finite unicity distance exists only if, for some
$n$, there is no \emph{redundant key use} in the cryptosystem,
i.e., no plaintext sequence $\mathbf{x}_n$ is mapped to the same
ciphertext $\mathbf{y}_n$ by more than one key value. With
redundant key use, one cannot pin down the key but it seems that
this may not enhance the system security either, and so is merely
wasteful. The exact possibilities will be analyzed elsewhere. For
now, we call a
cipher \emph{nondegenerate} 
in this paper if it has no redundant key use for some finite $n$
or for $n \rightarrow \infty$. 
Under the condition
\begin{equation} \label{nondegenerate}
\lim_{n \rightarrow \infty} H(\mathbf{Y}_n|\mathbf{X}_n) =
H(K),\end{equation} which is similar but not identical to the
definition of a `nondegenerate' cipher given in \cite{jkm}, one
may show that, when Eq.~(\ref{nonrandom}) also holds, one has
\begin{equation} \label{broken} \lim_{n \rightarrow \infty}
H(K|\mathbf{X}_n,\mathbf{Y}_n) = 0,
\end{equation}
so that the system is asymptotically broken under a
known-plaintext attack. More generally, for a nonrandom cipher, we
define a
 \emph{nondegeneracy distance} $n_d$ to be the smallest $n$ such that
 \begin{equation} \label{nondegdist}
 H(\mathbf{Y}_{n}|\mathbf{X}_{n})=H(K)
 \end{equation}
 holds, with $n_d =\infty$ if (\ref{nondegenerate}) holds and there is no finite $n$ satisfying
 (\ref{nondegdist}). Thus, a nonrandom cipher is nondegenerate in
 our sense if it has a nondegeneracy distance, finite or infinite. In general, of course, the cipher
 may be \emph{degenerate}, i.e., it has no nondegeneracy distance.
 We can readily show (see Appendix A of \cite{yuen05qph}) that, under known-plaintext attack, a nonrandom nondegenerate
cipher is broken at data length $n=n_d$, in the sense that
\begin{equation} \label{kpabroken}H(K|\mathbf{X}_{n_d}\mathbf{Y}_{n_d})=0.
\end{equation}

More generally, for both random and nonrandom ciphers, we define
the \emph{unicity distance under known-plaintext attacks}, denoted
by $n_1$, to be the smallest integer such that
\begin{equation} \label{unicitydistKPA}
H(K|\mathbf{X}_{n_1}\mathbf{Y}_{n_1}) = 0. \end{equation} If no
such integer exists, the unicity distance under KPA is taken to be
infinite if $\lim_{n \rightarrow \infty}
H(K|\mathbf{X}_n\mathbf{Y}_n) = 0$. Thus, $n_1$ is the minimum
length of data needed to break the cipher for \emph{any} possible
known-plaintext $\mathbf{X}_n$. For a nonrandom cipher, it is
equal to the nondegeneracy distance.

Many ciphers including the one-time pad and LFSRs (linear feedback
shift registers \cite{stinson}) have finite $n_d$. Similar to the
case of $n_d$ for nonrandom ciphers, $n_1$ for a random cipher may
not always exist. For our definition of $n_1$ to make sense for
random ciphers, we will impose a `nondegeneracy' restriction on
random ciphers: A \emph{random cipher} is said to be
\emph{nondegenerate} if and only if \emph{each} nonrandom cipher
resulting from an assignment $\mathbf{R}=\mathbf{r}$ of the
randomizer is nondegenerate. Then we say it has
\emph{information-theoretic security against known-plaintext
attacks} if
\begin{equation}  \label{ITsecurityKPA}
\inf_n H(K|\mathbf{X}_n,\mathbf{Y}_n) \neq 0,
\end{equation}
i.e., if $H(K|\mathbf{X}_n,\mathbf{Y}_n)$ cannot be made
arbitrarily small whatever $n$ is. In other words, $n_1$ does not
exist. The actual level of the information-theoretic security is
quantified by the left side of (\ref{ITsecurityKPA}). One major
motivation to study random ciphers is the \emph{possibility} that
they possess such information-theoretic security. Some discussion
on this point is also available in Appendix A of \cite{yuen05qph}.

Even in the absence of information-theoretic security,
nondegenerate random ciphers can be expected (see the discussion
in Section 2.2) to have larger unicity distance $n_1$ under KPA
compared to the case where the randomization is turned off. This
would, as assumed in cryptography practice, increase the
complexity of attacking the key significantly. If
Eq.~(\ref{kpabroken}) holds when $\mathbf{X}_{n}$ is replaced by a
specific $\mathbf{x}_{n}$, $n$ defines the unicity distance
corresponding to $\mathbf{x}_{n}$. The overall unicity distance
under KPA may be defined by
\begin{equation} \label{overallunicitydistance}
\bar{n}_1= \min_{H(K|\bf{X}_n = \bf{x}_n, Y_{n})=0} n \textup{ for
some } \bf{x}_n.
\end{equation}

The above result has not been given in the literature, perhaps
because $H(K|\mathbf{X}_n\mathbf{Y}_n)$ has not been used
previously to characterize known-plaintext attacks. Nevertheless,
it is assumed to be true in cryptography practice that $K$ would
be pinned down for sufficiently long $n$ in a nonrandom
`nondegenerate' cipher.

We now discuss the advantages that a random cipher provides as
compared to nonrandom ciphers. For the case of STA on the key when
the plaintext  $\mathbf{X}_n$ has nonuniform but i.i.d.
statistics, the so-called \emph{homophonic substitution} method
provides complete information-theoretic security, i.e.
$H(K|\mathbf{Y}_n)=H(K)$  \cite{jkm}.  The original form of
homophonic substitution involves assigning to each plaintext
symbol a number of possible \emph{sequences} of length $l$
proportional to its a priori probability in such a way that all
possible $l$-sequences are covered. Then, for every input symbol,
if one of its assigned $l$-sequences is generated at random, the
net effect is to generate $l$-sequences of plaintext with i.i.d.
uniform statistics. These sequences may be passed through a
non-degenerate cipher without revealing information on the key as
per Eq.~(\ref{keysecurity}). To put it another way, a statistical
attack has been converted to a ciphertext-only attack. A
generalized homophonic substitution that allows each symbol to be
coded into sequences of variable length is discussed in
\cite{jkm}, for which it is shown that sometimes data compression
instead of data expansion results.

Unfortunately, this reduction of a STA to a CTA does not work for
known-plaintext attacks. However, we emphasize that there is
\emph{no result} on random ciphers analogous to
Eq.~(\ref{kpabroken} ) with $n_d$ replaced by any definite $n$
depending on the cipher, since under randomization,
Eq.~(\ref{nonrandom}), and usually (\ref{nondegdist}) also, does
not hold for any $n$. Indeed, an inspection of the defining
equation Eq.~(\ref{Gammarandomcipherdef}) for a random cipher (or
Fig.~1) suggests how a random cipher may provide greater security
against KPAs. For a given plaintext-ciphertext sequence pair,
Eq.(\ref{Gammarandomcipherdef}) suggests that one has some
residual uncertainty on the value of the keystream $(Z_1,\ldots,
Z_n)$, which does not exist for a corresponding nonrandom cipher.
On the other hand, Eq.(\ref{Gammarandomcipherdef}) refers only to
the per-symbol uncertainty of the key stream calculated without
regard to the ciphertext observed for the other symbols in the
sequence. When such correlations are taken into account, the
uncertainty on the keystream may be drastically reduced and we can
give no general quantitative assertions of information-theoretic
security. Note, however, that due to the randomization, the
unicity distance $n_1$ of a random cipher under known-plaintext
attacks can be expected to be bigger than that of any of its
nonrandom reductions. Thus, the complexity-based security would be
greater.

In fact, the general problem of attacking a random cipher has
received limited attention because they are \emph{not used in
practice} due to the associated reduction in effective bandwidth
or data rate as is evident in homophonic substitution, due to the
need for high speed random number generation, and also due to the
uncertainty on the actual input statistics needed for, e.g.,
homophonic substitution randomization. Thus, the rigorous
quantitative security of symmetric-key random ciphers against
known-plaintext attacks is not known theoretically or empirically,
although in principle random ciphers have actual and potential
advantages just discussed.


\begin{thebibliography}{10}
\bibitem{note1}
The point that the throughput for $\alpha\eta$ key generation
using the optimal quantum receiver principle is low, which we
already remarked in \cite{yuen03,yuen05pla}, was also made in
\cite{nishioka04,nishioka05}. We would like to point out here that
the throughput can be increased to the level of the original data
rate by DSR. This involves moving the state to the basis boundary
with Bob utilizing a different matching quantum measurement. For
the case of key generation with uniform data, the output state is
still uniformly distributed on the circle on a per-qumode basis to
Eve. However, Bob's optimal performance has not been characterized
with DSR. A detailed discussion of $\alpha\eta$ key generation
will be presented later. We may observe that the use of DSR in
direct encrypion would necessitate the use of an error-correcting
mechanism in contrast with the original $\alpha\eta$, and would
break the Shannon limit (12). The significance and practicality of
such variations of $\alpha\eta$ have been briefly mentioned in
\cite{yuen03}, and will be treated elsewhere.
\bibitem{note2}
Note that for the purpose of bounding Eve's information in key
generation by granting her the key after her measurement, she
would make at least $0.1-1\%$ errors as asserted in
\cite{yuen05pla}.
\end{thebibliography}
\end{document}